%%%%%%%%%%%%%%%%%%%%%
%  lanlmac vs harvmac
%%%%%%%%%%%%%%%%%%%%%
\openin 1 lanlmac
\ifeof 1
  \message{[Load harvmac]}
  \input harvmac
\else
  \message{[Load lanlmac]}
  \input lanlmac
\fi
\closein 1

%%%%%%%%%%%%%%%%%%%%%%
%  packages
%%%%%%%%%%%%%%%%%%%%%%
\input amssym
%\draftmode
\noblackbox

%%%%%%%%%%%%%%%%%%%%%%
%  hypertex
%%%%%%%%%%%%%%%%%%%%%%
\newif\ifhypertex
\ifx\hyperdef\UnDeFiNeD
    \hypertexfalse
    \message{[HYPERTEX MODE OFF]}
    \def\hyperref#1#2#3#4{#4}
    \def\hyperdef#1#2#3#4{#4}
    \def\hypernoname{}
    \def\e@tf@ur#1{}
    
\else
    \hypertextrue
    \message{[HYPERTEX MODE ON]}
    
\fi

%%%%%%%%%%%%%%%%%%%%%%
%  Sizes
%%%%%%%%%%%%%%%%%%%%%%

\baselineskip=16pt plus 2pt minus 1pt
\parskip=2pt plus 16pt minus 1pt

%%%%%%%%%%%%%%%%%%%%%%
%  Tables
%%%%%%%%%%%%%%%%%%%%%%
\newcount\tabno
\tabno=1
\def\ltab#1{%
\let\labelflag=#1%
\def\numb@rone{#1}%
\ifx\labelflag\UnDeFiNeD{%
  \xdef#1{\the\tabno}%
  \writedef{#1\leftbracket{\the\tabno}}%
  \global\advance\tabno by1%
}%
\fi%
{\hyperref{}{table}{{\numb@rone}}{Table~{\numb@rone}}}}
\def\tabinsert#1#2#3{%
\let\flag=#1
\ifx\flag\UnDeFiNeD
  {\xdef#1{\the\tabno}
   \writedef{#1\leftbracket{\the\tabno}}
   \global\advance\tabno by1 }
\fi
\vbox{\bigskip #3 \smallskip
\leftskip 4pc \rightskip 4pc
\noindent\ninepoint\sl \baselineskip=11pt
{\bf{\hyperdef\hypernoname{table}{{#1}}{Table~{#1}}}.~}#2
\smallskip}
\bigskip}

%%%%%%%%%%%%%%%%%%%%%%
%  macros
%%%%%%%%%%%%%%%%%%%%%%

\def\cf{{\it cf.\ }}
% 'double' letters

\def\IP{{\Bbb P}}

\def\IW{{\Bbb W}}
\def\IZ{{\Bbb Z}}
% calligraphic letters

\def\var1{z}
%%%%%%%%%%%%%%%%%%%%%%
\def\frac#1#2{{#1\over#2}}
\def\rel#1{\underline{#1}}
\def\HGF#1#2{{{}_#1F_#2}}

%\def\HJ#1{{\bf HJ: #1}}
%\def\MS#1{{\bf MS: #1}}

%%%%%%%%%%%%%%%%%%%%%%
%  references
%%%%%%%%%%%%%%%%%%%%%%

%\AganagicGS
\lref\AganagicGS{
M.~Aganagic and C.~Vafa,
``Mirror symmetry, D-branes and counting holomorphic discs,''
arXiv:hep-th/0012041.
%%CITATION = HEP-TH/0012041;%%
}

%\AganagicNX
\lref\AganagicNX{
M.~Aganagic, A.~Klemm and C.~Vafa,
``Disk instantons, mirror symmetry and the duality web,''
Z.\ Naturforsch.\  A {\bf 57}, 1 (2002)
[arXiv:hep-th/0105045].
%%CITATION = ZNTFA,A57,1;%%
}

%\AlimRF
\lref\AlimRF{
M.~Alim, M.~Hecht, P.~Mayr and A.~Mertens,
``Mirror Symmetry for Toric Branes on Compact Hypersurfaces,''
arXiv:0901.2937 [hep-th].
%%CITATION = ARXIV:0901.2937;%%
}

%\BerglundPW
\lref\BerglundPW{
P.~Berglund and S.~H.~Katz,
``Mirror symmetry for hypersurfaces in weighted projective space and
topological couplings,''
Nucl.\ Phys.\  B {\bf 420}, 289 (1994)
[arXiv:hep-th/9311014].
%%CITATION = NUPHA,B420,289;%%
}

%\CandelasRM
\lref\CandelasRM{
P.~Candelas, X.~C.~De La Ossa, P.~S.~Green and L.~Parkes,
``A pair of Calabi-Yau manifolds as an exactly soluble superconformal
theory,''
Nucl.\ Phys.\  B {\bf 359}, 21 (1991).
%%CITATION = NUPHA,B359,21;%%
}

%\CandelasDM
\lref\CandelasDM{
P.~Candelas, X.~De La Ossa, A.~Font, S.~H.~Katz and D.~R.~Morrison,
``Mirror symmetry for two parameter models. I,''
Nucl.\ Phys.\  B {\bf 416}, 481 (1994)
[arXiv:hep-th/9308083].
%%CITATION = NUPHA,B416,481;%%
}

%\CandelasHW
\lref\CandelasHW{
P.~Candelas, A.~Font, S.~H.~Katz and D.~R.~Morrison,
``Mirror symmetry for two parameter models. 2,''
Nucl.\ Phys.\  B {\bf 429}, 626 (1994)
[arXiv:hep-th/9403187].
%%CITATION = NUPHA,B429,626;%%
}

%\CandelasRM
\lref\CandelasRM{
P.~Candelas, X.~C.~De La Ossa, P.~S.~Green and L.~Parkes,
``A pair of Calabi-Yau manifolds as an exactly soluble superconformal
theory,''
Nucl.\ Phys.\  B {\bf 359}, 21 (1991).
%%CITATION = NUPHA,B359,21;%%
}

%\GreeneUD
\lref\GreeneUD{
B.~R.~Greene and M.~R.~Plesser,
``Duality in Calabi-Yau moduli space,''
Nucl.\ Phys.\  B {\bf 338}, 15 (1990).
%%CITATION = NUPHA,B338,15;%%
}

%\GrimmDQ
\lref\GrimmDQ{
T.~W.~Grimm, T.~W.~Ha, A.~Klemm and D.~Klevers,
``The D5-brane effective action and superpotential in N=1
compactifications,''
arXiv:0811.2996 [hep-th].
%%CITATION = ARXIV:0811.2996;%%
}

%\JockersPE
\lref\JockersPE{
H.~Jockers and M.~Soroush,
``Effective superpotentials for compact D5-brane Calabi-Yau geometries,''
arXiv:0808.0761 [hep-th].
%%CITATION = ARXIV:0808.0761;%%
}

%\KachruIH
\lref\KachruIH{
S.~Kachru, S.~H.~Katz, A.~E.~Lawrence and J.~McGreevy,
``Open string instantons and superpotentials,''
Phys.\ Rev.\  D {\bf 62}, 026001 (2000)
[arXiv:hep-th/9912151].
%%CITATION = PHRVA,D62,026001;%%
}

%\KreflSJ
\lref\KreflSJ{
D.~Krefl and J.~Walcher,
``Real Mirror Symmetry for One-parameter Hypersurfaces,''
JHEP {\bf 0809}, 031 (2008)
[arXiv:0805.0792 [hep-th]].
%%CITATION = JHEPA,0809,031;%%
}

%\LabastidaZP
\lref\LabastidaZP{
J.~M.~F.~Labastida and M.~Mari\~no,
``Polynomial invariants for torus knots and topological strings,''
Commun.\ Math.\ Phys.\  {\bf 217}, 423 (2001)
[arXiv:hep-th/0004196].
%%CITATION = CMPHA,217,423;%%
}

%\LercheCK
\lref\LercheCK{
W.~Lerche, P.~Mayr and N.~Warner,
``Holomorphic N = 1 special geometry of open-closed type II strings,''
arXiv:hep-th/0207259.
%%CITATION = HEP-TH/0207259;%%
}

%\LercheCW
\lref\LercheCW{
W.~Lerche and P.~Mayr,
``On N = 1 mirror symmetry for open type II strings,''
arXiv:hep-th/0111113.
%%CITATION = HEP-TH/0111113;%%
}

%\LercheYW
\lref\LercheYW{
W.~Lerche, P.~Mayr and N.~Warner,
``N = 1 special geometry, mixed Hodge variations and toric geometry,''
arXiv:hep-th/0208039.
%%CITATION = HEP-TH/0208039;%%
}

%\MayrSH
\lref\MayrSH{
P.~Mayr,
``Mirror symmetry, N = 1 superpotentials and tensionless strings on
Calabi-Yau four-folds,''
Nucl.\ Phys.\  B {\bf 494}, 489 (1997)
[arXiv:hep-th/9610162].
%%CITATION = NUPHA,B494,489;%%
}

%\MayrXK
\lref\MayrXK{
P.~Mayr,
``N = 1 mirror symmetry and open/closed string duality,''
Adv.\ Theor.\ Math.\ Phys.\  {\bf 5}, 213 (2002)
[arXiv:hep-th/0108229].
%%CITATION = 00203,5,213;%%
}

%\MorrisonBM
\lref\MorrisonBM{
D.~R.~Morrison and J.~Walcher,
``D-branes and Normal Functions,''
arXiv:0709.4028 [hep-th].
%%CITATION = ARXIV:0709.4028;%%
}

%\OoguriBV
\lref\OoguriBV{
H.~Ooguri and C.~Vafa,
``Knot invariants and topological strings,''
Nucl.\ Phys.\  B {\bf 577}, 419 (2000)
[arXiv:hep-th/9912123].
%%CITATION = NUPHA,B577,419;%%
}

%\OoguriCK
\lref\OoguriCK{
H.~Ooguri, Y.~Oz and Z.~Yin,
``D-branes on Calabi-Yau spaces and their mirrors,''
Nucl.\ Phys.\  B {\bf 477}, 407 (1996)
[arXiv:hep-th/9606112].
%%CITATION = NUPHA,B477,407;%%
}

%\KnappUW
\lref\KnappUW{
J.~Knapp and E.~Scheidegger,
``Towards Open String Mirror Symmetry for One-Parameter Calabi-Yau
Hypersurfaces,''
arXiv:0805.1013 [hep-th].
%%CITATION = ARXIV:0805.1013;%%
}

%\SchwarzTI
\lref\SchwarzTI{
A.~Schwarz and V.~Vologodsky,
``Integrality theorems in the theory of topological strings,''
arXiv:0807.1714 [hep-th].
%%CITATION = ARXIV:0807.1714;%%
}

%\WalcherRS
\lref\WalcherRS{
J.~Walcher,
``Opening mirror symmetry on the quintic,''
Commun.\ Math.\ Phys.\  {\bf 276}, 671 (2007)
[arXiv:hep-th/0605162].
%%CITATION = CMPHA,276,671;%%
}

%\WittenFB
\lref\WittenFB{
E.~Witten,
``Chern-Simons Gauge Theory As A String Theory,''
Prog.\ Math.\  {\bf 133}, 637 (1995)
[arXiv:hep-th/9207094].
%%CITATION = PMTMA,133,637;%%
}

%\WittenYC
\lref\WittenYC{
E.~Witten,
``Phases of N = 2 theories in two dimensions,''
Nucl.\ Phys.\  B {\bf 403}, 159 (1993)
[arXiv:hep-th/9301042].
%%CITATION = NUPHA,B403,159;%%
}

\lref\WProg{
M.~Alim, M.~Hecht, H.~Jockers, P.~Mayr, A.~Mertens and M.~Soroush,
work in progress.}

%%%%%%%%%%%%%%%%%%%%%%
%  end references
%%%%%%%%%%%%%%%%%%%%%%

%%%%%%%%%%%%%%%   Title Page  %%%%%%%%%%%%%
\Title{\vbox{\hbox{\tt SU-ITP-09/18}} }
{\vbox{
  \centerline{\hbox{Relative periods and open-string integer invariants}}
  \vskip 0.7cm
  \centerline{\hbox{for a compact Calabi-Yau hypersurface}}
}}
\bigskip
\centerline{\bf Hans Jockers\footnote{${}^{\,a}$}{\tt jockers@stanford.edu}
                       and Masoud Soroush\footnote{${}^{\,b}$}{\tt soroush@stanford.edu}}
\bigskip
\centerline{{\it Department of Physics, Stanford University}}
\centerline{{\it  Stanford, CA 94305-4060, USA}}
\bigskip

\vskip 0.5cm
\centerline{\bf Abstract}
In this work we compute relative periods for B-branes, realized in terms of divisors in a compact Calabi-Yau hypersurface, by means of direct integration. Although we exemplify the method of direct integration with a particular Calabi-Yau geometry, the recipe automatically generalizes for divisors in other Calabi-Yau geometries as well. From the calculated relative periods we extract double-logarithmic periods. These periods qualify to describe disk instanton generated $N=1$ superpotentials of the corresponding compact mirror Calabi-Yau geometry in the large volume regime. Finally we extract the integer invariants encoded in these brane superpotentials.

\bigskip
\Date{\sl {April, 2009}}
\vfill\eject

%%%%%%%%%%%%%%%%%%%%%%%%%%%%%%%%%%
\newsec{Introduction}
%%%%%%%%%%%%%%%%%%%%%%%%%%%%%%%%%%
The study of mirror symmetry has always revealed many interesting results. This duality has triggered the systematic development of the notion of quantum geometry in string theory, not only on a conceptional level but also from a computational point of view. In practice mirror symmetry often allows us to compute certain worldsheet instanton corrected couplings, which are not easily accessible by other techniques. In addition to its applications in physics the discovery of mirror symmetry has strongly influenced the development of enumerative geometry in mathematics.

In the context of mirror symmetry holomorphic quantities in the effective description of Calabi-Yau string compactifications are most easily accessible. For instance, in the context of $N=2$ string compactifications the holomorphic prepotential can be computed in the topological B-model by means of classical geometry, which, then by mirror symmetry is equivalent to the worldsheet instanton generated quantum prepotential of the dual mirror string compactification. These instanton corrections correspond to closed-string Gromov-Witten invariants. Moreover, the mirror symmetry recipe has also been generalized to calculate open-string worldsheet instantons in the presence of D-branes \refs{\AganagicGS,\AganagicNX,\LercheCK,\LercheYW}. The generating functions of these disk instanton corrections describe certain quantum generated $N=1$ brane superpotentials in the effective description of Calabi-Yau compactifications with D-branes \refs{\WittenFB,\OoguriCK,\KachruIH}. From the perspective of enumerative geometric the open-string Gromov-Witten invariants are not as well understood as their closed string ancestors.    

So far, most explicit results for open-string disk instantons deal with branes in local, that is to say, non-compact target space geometries. Recently, however, there has been significant progress in computing disk instantons in the context of compact Calabi-Yau geometries. In a series of remarkable papers \refs{\WalcherRS,\MorrisonBM,\KreflSJ,\KnappUW}, the authors compute disk instanton invariants for involution branes in compact Calabi-Yau geometries. Physically, the computed disk instanton generating functions capture the domain wall tension between supersymmetric vacua of two distinct D-brane configurations.

In refs.~\refs{\MayrXK,\LercheCW,\LercheCK,\LercheYW} the authors derive and analyze the structure of Picard-Fuchs differential equations governing the relative periods of brane geometries in non-compact threefolds. Recently, these techniques have also been generalized to capture $N=1$ superpotentials for compact Calabi-Yau geometries \JockersPE. Furthermore, it is demonstrated that by evaluating these $N=1$ superpotentials at their critical points, the previously mentioned instanton generated domain wall tensions are reproduced. In ref.~\AlimRF\ this approach is further extended and refined. For the first time in the large radius regime the authors manage to extract disk instantons arising from D-branes with a classically unobstructed open-string modulus in the context of compact Calabi-Yau threefolds. The encoded open-string Gromov-Witten invariants agree in suitable decompactification limits with previously established computations for branes in non-compact Calabi-Yau geometries \AganagicNX.

Although there has recently been much progress in revealing the structure of brane superpotentials in compact geometries, many aspects are still unclear and need to be studied in greater detail. The goal of this note is to present the technique of direct integration to derive the relative periods in the topological B-model, which then serve as a starting point to analyze the underlying quantum geometry of the associated A-model geometry. We demonstrate the procedure of direct integration for a particular class of divisors in compact Calabi-Yau geometries on the B-model side. For the bulk geometry the procedure of direct integration has previously been developed in refs.~\refs{\CandelasRM,\CandelasDM,\CandelasHW,\BerglundPW}.

Instead of describing the procedure in full generality we exemplify our approach by explicitly studying two different divisors in the mirror of the Calabi-Yau hypersurface of the weighted projective space, $\IW\IP^4_{(1,1,2,2,6)}$ (\cf ref.~\CandelasDM). However, the presented method works more generally, and it is straight forward to apply it to other Calabi-Yau geometries as well.

Finally, we use the derived relative periods to extract generating functions for integral invariants. We believe that the integrality of the stated invariants is a general feature \SchwarzTI, and we argue that the computed generating functions are natural candidates for disk instanton generated $N=1$ superpotentials for the associated A-model configuration, which is the mirror to our discussed B-model setup. However, in order to really establish this correspondence further checks have to be performed, which are beyond the scope of this work. Namely, it is necessary to study in detail the systematics of the structure of relative periods in the B-model so as to explicitly construct the associated A-model geometry \WProg.

The outline of this paper is as follows: In Section~2 we briefly introduce the considered Calabi-Yau and brane geometry. Then in Section~3 we demonstrate with the given geometry the technique of direct integration to derive the relative periods, which are relevant to extract $N=1$ brane superpotentials. In Section~4 we use the computed results to extract integer invariants, which contingently give rise to disk instanton generated $N=1$ brane superpotentials for an appropriate mirror A-model brane configuration. Finally we present in Section~5 our conclusions.

%%%%%%%%%%%%%%%%%%%%%%
\newsec{The B-model Geometry}
%%%%%%%%%%%%%%%%%%%%%%
In this section we introduce the bulk and D-brane configurations of the analyzed B-model geometry. The bulk Calabi-Yau geometry is given in terms of the mirror manifold of a hypersurface in a weighted projective space. This mirror pair is analyzed in detail in ref.~\CandelasDM. Analogously to refs.~\refs{\AganagicGS,\LercheCK,\LercheYW,\JockersPE,\AlimRF} the brane geometry of the B-model is realized as a divisor in this Calabi-Yau space.

%%%%%%%%%%%%%%%%%%%%%%
\subsec{The bulk geometry}
%%%%%%%%%%%%%%%%%%%%%%
The relevant mirror Calabi-Yau threefold pair has been studied in detail in ref.~\CandelasDM. The A-model bulk Calabi-Yau geometry, $X$, is given by the family of degree twelve hypersurfaces in the weighted projective space, $\IW\IP^4_{(1,1,2,2,6)}$, parametrized by the weighted projective coordinates $[\,x_1\,:\,x_2\,:\,x_3\,:\,x_4\,:x_5\,]$. These hypersurfaces have a genus two curve of $A_1$~singularities along the locus $x_1=x_2=0$. Resolving these singularities we obtain a Calabi-Yau hypersurface with two K\"ahler and $128$ complex structure moduli. The K\"ahler cone is spanned by a line in the linear system of degree one polynomials and by a fiber of the resolution of the curve of $A_1$~singularities \CandelasDM.

The mirror family of Calabi-Yau geometries, $Y$, can be obtained by the Greene-Plesser construction \GreeneUD, and we choose the degree twelve hypersurface polynomial,
\eqn\HyperPol{
  P(\psi,\phi)\,=\,x_1^{12}+x_2^{12}+x_3^6+x_4^6+x_5^2-12\,\psi\,x_1x_2x_3x_4x_5-2\,\phi\,x_1^6x_2^6 \ , }
in the weighted projective space, $\IW\IP^4_{(1,1,2,2,6)}$, modded out by the Greene-Plesser group, $\IZ_6^2\times\IZ_2$, generated by
\eqn\GPgroup{g_1\,=\,(0,5,1,0,0) \ , \quad g_2\,=\,(0,5,0,1,0) \ , \quad g_3\,=\,(0,1,0,0,1) \ , }
For instance the generator, $g_1$, acts on the coordinates, $[x_1\,:\,x_2\,:\,x_3\,:\,x_4\,:\,x_5]$, of the weighted projective space as
\eqn\Gaction{g_1\,:\  [x_1\,:\,x_2\,:\,x_3\,:\,x_4\,:\,x_5] \mapsto [x_1\,:\,\eta^5 x_2\,:\,\eta x_3\,:\,x_4\,:\,x_5] \ ,
  \quad \eta\equiv e^{2\pi i/6} \ . }
By standard techniques the resulting singular Calabi-Yau threefolds, $Y$, can be resolved to a family of smooth Calabi-Yau threefolds with $128$ K\"ahler and two complex structure moduli. The latter moduli are parametrized by the algebraic coordinates, $\psi$ and $\phi$,  in the hypersurface polynomial~\HyperPol.

%%%%%%%%%%%%%%%%%%%%%%
\subsec{The B-brane geometry}
\subseclab\secBbraneGeom
%%%%%%%%%%%%%%%%%%%%%%
As in refs.~\refs{\JockersPE,\AlimRF} we capture the brane geometry in terms of additional homogenous equations, which correspond to divisors of the Calabi-Yau manifold, $Y$. The two families of branes analyzed in this work are given in terms of the two divisors
\eqn\DivisorQOne{Q_1(\xi)\,=\,x_5-\xi\,x_1x_2x_3x_4 \ , }
and
\eqn\DivisorQTwo{Q_2(\zeta)\,=\,x_1^6-\zeta\,x_2^6 \ . }
Here the variables, $\xi$ and $\zeta$, represent the algebraic open-string moduli of the two families of B-brane geometries respectively. In order to be compatible with the resolved bulk geometry, the divisors are required to have definite charges with respect to the Greene-Plesser group~$\IZ_6\times\IZ_6\times\IZ_2$. Note that both divisors, $Q_1$ and $Q_2$, are the most general degree six polynomials with charges $(0,0,1)$ and $(0,0,0)$, respectively.

In the presence of the first divisor, $Q_1$, the open-closed geometry exhibits an enhanced discrete symmetry. The algebraic variables, $\psi$ and $\phi$, of the bulk hypersurface polynomial, $P(\psi,\phi)$, and the algebraic variable, $\xi$, of the divisor polynomial, $Q(\xi)$, fulfill a discrete $\IZ_{12}\times\IZ_2$~symmetry, which is generated by
\eqn\SymAlgQOne{\eqalign{
  \IZ_{12}\,:&\ (\psi,\phi,\xi)\,\mapsto\,(\rho\,\psi,-\phi,\rho\,\xi) \ , \quad \rho\,\equiv\,e^{2\pi i/12} \ , \cr
  \IZ_2\,:&\ (\psi,\phi,\xi)\,\mapsto\,(\psi,\phi,\xi-12\psi) \ . }}
This discrete symmetry action can easily be checked by acting on the homogeneous polynomials, $P$ and $Q_1$, together with a simple change of the projective coordinates. Similarly we find for the second divisor, $Q_2$, the discrete $\IZ_{12}$~symmetry
\eqn\SymAlgQTwo{
  \IZ_{12}\,: \ (\psi,\phi,\zeta)\,\mapsto\,(\rho\,\psi,-\phi,-\zeta) \ , \quad \rho\,\equiv\,e^{2\pi i/12} \ . }
%

%%%%%%%%%%%%%%%%%%%%%%
\subsec{The A-brane geometry}
%%%%%%%%%%%%%%%%%%%%%%
Let us briefly turn to the mirror A-brane configurations associated to the introduced B-branes. Applying the techniques introduced in refs.~\refs{\AganagicGS,\AlimRF}, we find that the A-brane geometry is constrained by the gauged linear $\sigma$-model charge vector\foot{For more detail we refer to ref.~\AlimRF.}
\eqn\CVecOne{ \hat l^1 \,=\,( -1 , 0 , 0 ,0 ,0, 1) \ , }
which give also rise to the constraint
\eqn\DivisorQOne{ -|p|^2+|x_5|^2\,=\,c \ . }
This equation constraints the family of A-branes, that is to say the family of special Lagrangian submanifolds, in the gauged linear $\sigma$-model of the Calabi-Yau hypersurface. The expectation value of the field, $p$, parametrizes (one of) the K\"ahler moduli, and it acquires the expectation value, $p=0$, in the large radius regime \WittenYC. In this geometric semi-classical phase the fields, $x_\ell$, become the projective coordinates of the ambient weighted projective space, $\IW\IP^4_{(1,1,2,2,6)}$. The parameter, $c$, encodes the position of the A-brane, and the semi-classical brane configuration, in which worldsheet instanton corrections are suppressed, is obtained for large values of the parameter, $c$. For more details we refer the reader to refs.~\refs{\AganagicGS,\AlimRF}.

Note, however, that the A-brane geometry is not solely determined by the constrained~\DivisorQOne. This is due to the fact that also on the B-model side the B-brane geometry is not completely specified by the divisor~\DivisorQOne. As discussed in detail in Section~4 the B-brane configuration is further constrained by specifying the brane superpotential as a linear combination of (relative) periods. Therefore, similarly as for the examples in ref.~\AlimRF, on the mirror side the A-brane geometry must also be further constraint. We come back to this issue elsewhere \WProg.

For the second divisor, $Q_2$, the gauged linear $\sigma$-model mirror symmetry construction yields the A-brane charges
\eqn\CVecTwo{ \hat l^1 \,=\,( 0 , 1 , -1 ,0, 0, 0) \ . }
and relation
\eqn\DivisorQTwoLR{ |x_1|^2-|x_2|^2\,=\,c \ . }
Analogously as in the previous example the parameter, $c$, determines the position of the A-brane, and the semi-classical phase arises for large value of $c$.

%%%%%%%%%%%%%%%%%%%%%%
\newsec{Relative periods and the mirror map}
%%%%%%%%%%%%%%%%%%%%%%
The differential equations for the relative periods are governed by the variation of mixed Hodge structure and they can be explicitly computed by deriving the Picard-Fuchs equations from residue integrals of the relative three-forms. This procedure has been carried out for the non-compact Calabi-Yau geometries in refs.~\refs{\AganagicGS,\LercheCK,\LercheYW} and for compact geometries in ref.~\JockersPE. For instance for the compact Calabi-Yau threefold, $Y$, with the divisor, $Q_1$, we realize the unique relative three-form, $\rel\Omega$, in terms of the residue integral
\eqn\ResThreeForm{\rel{\Omega}(\psi,\phi,\xi)\,\sim\,\int\frac{\log Q_1(\xi)}{P(\psi,\phi)}\,\Delta_x \ , }
with the differential
\eqn\Diff{
  \Delta_x\,=\,\sum_{n=1}^5(-1)^na_n x_n dx_1\wedge\ldots\widehat{dx_n}\ldots \wedge dx_5 \ , }
where the integers, $a_n$, denote the weights of the weighted projective space, $\IW\IP^5_{(1,1,2,2,6)}$. As explained in detail in ref.~\JockersPE\ by taking derivates of the relative three-form~$\rel\Omega$ with respect to the algebraic parameters, $\psi$, $\phi$ and $\xi$, we obtain a system of linear differential equations, which can eventually be rewritten to Picard-Fuchs differential equations governing the relative periods.

For toric branes in compact toric Calabi-Yau spaces the derivation of the GKZ system has systematically been described in ref.~\AlimRF. The presented techniques allow to compute the GKZ system elegantly and economically. The resulting relevant relative periods obtained from the Picard-Fuchs system and the GKZ system are equivalent, and the relationship between these two approaches are discussed in detail in the upcoming work \WProg.

In ref.~\GrimmDQ\ yet another alternative procedure is proposed to compute differential equations governing the open-/closed-string deformation space. It would be interesting to verify this proposal explicitly and to also make the connection to the techniques presented here and in refs.~\refs{\JockersPE,\AlimRF}.

%%%%%%%%%%%%%%%%%%%%%%
\subsec{The method of direct integration}
%%%%%%%%%%%%%%%%%%%%%%
Instead of extracting the differential equations from the residue integrals~\ResThreeForm, we compute the relevant relative periods by direct integration \refs{\CandelasRM,\CandelasDM}.

Let us first briefly review the idea of direct integration for the fundamental period of the bulk Calabi-Yau geometry, $Y$. By choosing an appropriate three cycle, $\Gamma$, the fundamental period is given by\foot{For later convenience we have introduced the pre-factor $-\frac{12\psi}{(2\pi i)^4}$ as a normalization for the fundamental period.}
\eqn\FundPeriod{\Pi_0(\psi,\phi)\,=\, -\frac{12\psi}{(2\pi i)^4} \int_\Gamma \Omega(\psi,\phi) \ , }
which, after inserting for the three form, $\Omega$, the residue integral representation, becomes the five-fold residue integral in the large complex structure regime \CandelasDM
\eqn\FundResiduePeriod{\eqalign{
  \Pi_0(\psi,\phi)\,&=\,-\frac{12\psi}{(2\pi i)^5}\int_{\gamma_1\times\ldots\times\gamma_5}\!\!\!\!\!
    \frac{dx_1dx_2dx_3dx_4dx_5}{P(\psi,\phi)}\cr
    \,&=\,\frac{1}{(2\pi i)^5}\int_{\gamma_1\times\ldots\times\gamma_5}\!\!\!\!\!
    \frac{dx_1dx_2dx_3dx_4dx_5}{x_1x_2x_3x_4x_5}\sum_{m=0}^{+\infty}
    \frac{(x_1^{12}+x_2^{12}+x_3^6+x_4^6+x_5^2-2\phi x_1^6x_2^6)^m}{(12\psi)^m(x_1x_2x_3x_4x_5)^m}\ . }}
Here the contours, $\gamma_\ell$, encircle the complex coordinates, $x_\ell$. Evaluating this residue integral reduces to a straight forward combinatorial problem, which yields for the fundamental period
\eqn\BulkFundPer{
  \Pi_0(\psi,\phi)\,=\,\sum_{n=0}^{+\infty}\frac{(6n)!(-1)^n}{(n!)^3(3n)!(12\psi)^{6n}}u_n(\phi) \ ,}
with
\eqn\UFunction{u_n(\phi)\,=\,(2\phi)^n\sum_{r=0}^{\left[\frac n2\right]}\frac{n!}{(r!)^2(n-2r)!(2\phi)^{2r}} \ . }
The general structure of the open-/closed differential Picard-Fuchs equations ensures that the closed string periods furnish relative periods of the open-/closed string deformation space. Hence we readily identify the fundamental relative period to be
\eqn\RelFundPer{\rel\Pi_0(\psi,\phi)\,=\,\Pi_0(\psi,\phi) \ . }

Let us now apply the techniques of direct integration to the subsystem of the open-/closed-string deformation space. The subsystem of relative forms is spanned by the relative three-forms, which are obtained as derivatives with respect to the algebraic open-string parameters \JockersPE. For divisors with a single open-string coordinate, which applies to the divisors~\DivisorQOne\ and \DivisorQTwo\ discussed in this work, the subsystem is particularly simple, as it is spanned by a single relative form and its derivatives. For the divisor, $Q_1$, this generating relative three-form is given by
\eqn\ResTwoForm{\partial_\xi\rel{\Omega}\,
   \simeq\,-\int\frac{x_1x_2x_3x_4}{P(\psi,\phi)\,Q_1(\xi)}\,\Delta_x \ .}

In order to make contact with the direct integration techniques reviewed for the fundamental period in the bulk, we first perform a change of coordiates given by
\eqn\ChangeVar{y_1=x_1^2\ ,\quad y_2=x_2^2\ , \quad y_3=x_3^2\ , \quad y_4=x_4^2\ ,\quad y_5=x_5-\xi\,x_1x_2x_3x_4 \ . }
Then the relative three-form~\ResTwoForm\ becomes
\eqn\ChangedResTwoForm{
  \partial_\xi\rel{\Omega}\,\sim\,\int \frac{1+O(y_5)}{(\hat P_1(\alpha,\phi)+O(y_5))\,y_5}\Delta_y\,dy_5 \ , }
in terms of the differential
\eqn\DiffFour{\Delta_y\,=\,\sum_{n=1}^4(-1)^n \hat a_n y_n dy_1\wedge\ldots\widehat{dy_n}\ldots \wedge dy_4 \ , \quad
\hat a_\ell\,=\,(1,1,2,2,6) \ , }
and the polynomial
\eqn\HyperPolSub{
    \hat P_1(\alpha,\phi)\,=\, y_1^6+y_2^6+y_3^3+y_4^3-\alpha\,y_1y_2y_3y_4-2\phi\,y_1^3y_2^3 \ , }
with the parameter
\eqn\AlphaPar{\alpha\,=\,\xi(12\,\psi-\xi) \ . }

By performing this change of variables we have transformed the residue integral~\ResTwoForm\ in the weighted projective space, $\IW\IP^4_{(1,1,2,2,6)}$, into a new residue integral in the weighted projective space, $\IW\IP^4_{(1,1,2,2,3)}$, parametrized by the projective coordinates, $[\, y_1\,:\,y_2\,:\,y_3\,:\,y_4\,:\,y_5\,]$. The obtained expression can be further simplified by performing the integral along the contour encircling the coordinate, $y_5$, and we arrive at
\eqn\ResTwoFormCY{\partial_\xi\rel\Omega\,\sim\,\int\frac{1}{\hat P_1(\alpha,\phi)}\Delta_y \ . }

We recognize the resulting residue integral as the unique holomorphic two-form of the two-moduli K3~hypersurface, $\hat P_1(\alpha,\phi)$, in the ambient projective space, $\IW\IP^3_{(1,1,2,2)}$, together with the induced Greene-Plesser group, $\IZ_3^2$, generated by
\eqn\SubOneGP{ \hat g_1\,=\,(0,2,1,0) \ , \quad \hat g_2\,=\,(0,2,0,1) \ . }
Thus, as a consequence the periods of the subsystem coincide with the periods of a certain K3~surface \AlimRF.

Moreover, we also observe that, although the subsystem geometry is associated to a single open-string parameter, $\xi$, the obtained subsystem K3~period depends on two algebraic moduli, $\alpha$ and $\phi$. As a consequence, as we will see, the geometry associated to the divisor, $Q_1$, encodes two superpotential candidates and, hence, encodes potentially two classes of distinct brane geometries.

Analogously to the bulk geometry we determine the fundamental period, $\chi_0(\alpha,\phi)$, of the subsystem again by the method of direct integration. This is achieved by evaluating the residue integral
\eqn\SubFundResiduePeriod{
  \chi_0(\alpha,\phi)\,=\,-\frac{12\psi}{(2\pi i)^4}\int_{\gamma_1\times\ldots\times\gamma_4}\!\!\!\!\!
    \frac{dy_1dy_2dy_3dy_4}{\hat P_1(\alpha,\phi)} \ , }
where, in order to be compatible with the fundamental period, $\rel\Pi_0$, we have included the same normalization factor $\psi$. Carrying out the residue integrals, we finally arrive at the subsystem fundamental period
\eqn\SubfundPerOne{
  \chi_0(\alpha,\phi)\,=\,\frac{\psi}{\alpha} \sum_n^{+\infty}\frac{(3n)!(-1)^n}{(n!)^3\alpha^{3n}}u_n(\phi) \ , }
together with eq.~\UFunction.

For the second divisor, $Q_2$, under consideration we proceed analogously. After the change of projective variables
\eqn\ChangeVarTwo{ y_0\,=\,x_1^6-\zeta x_2^6 \ , \quad y_1\,=\,(\zeta^2-2\phi\,\zeta+1)^\frac{1}{6}x_2^2 \ ,
  \quad y_2\,=\,x_3 \ , \quad y_3\,=\,x_4 \ , \quad y_4\,=\,x_5 \ . }
we arrive after a few steps of algebra and after integrating out the variable, $y_0$, at the one-modulus K3~subsystem residue integral\foot{Note that the prefactor, $\frac{\beta}{\zeta}$, arises from the Jacobian induced from change of variables~\ChangeVarTwo.}
\eqn\ResTwoFormCYTwo{\partial_\zeta\rel\Omega\,\sim\, \frac{12\psi\beta}{\zeta} \int\frac{1}{\hat P_2(\beta)}\Delta_y \ , }
where we have again multiplied by the variable, $12\psi$, in order to be in agreement with the chosen normalization of the relative fundamental period, $\rel\Pi_0$. The homogeneous polynomial, $\hat P_2(\beta)$, is given by
\eqn\HyperPolSubTwo{\hat P_2(\beta)\,=\, y_1^6+y_2^6+y_3^6+y_4^2-\beta\,y_1y_2y_3y_4 \ ,  \quad
   \beta\,=\,12\psi\left(\frac{\zeta}{\zeta^2-2\phi\,\zeta+1}\right)^\frac{1}{6} \ , }
in the weighted projective space, $\IW\IP^3_{(1,1,1,3)}$. For this divisor the representation of induced Greene-Plesser group is actually enlarged due to integrating out the $y_0$-coordinate. As a consequence the algebraic modulus, $\beta$, acquires also a discrete charge. For the enlarged generators of the induced Greene-Plesser group, $\IZ_6^2\times\IZ_2$, acting on the variables, $(y_1, y_2, y_3, y_4, \beta)$, we obtain
\eqn\SubTwoGP{ \hat g_1\,=\,(4,1,0,0,1) \ , \quad \hat g_2\,=\,(4,0,1,0,1)\ , \quad \hat g_3\,=\,(0,0,0,1,1) \ . }
Finally, extracting from eq.~\ResTwoFormCYTwo\ the fundamental period of the subsystem, $\vartheta_0$, via direct integration, we arrive at the hypergeometric function
\eqn\SubfundPerTwo{
  \vartheta_0(\beta)\,=\,-\frac{12\psi\beta}{(2\pi i)^4\zeta}\int_{\gamma_1\times\ldots\times\gamma_4}\!\!\!\!\!
    \frac{dy_1dy_2dy_3dy_4}{\hat P_2(\beta)}\,=\,\frac{1}{\zeta}\,\HGF{3}{2}\left(\frac{1}{6},\frac{3}{6},\frac{5}{6}; 1,1; \frac{1}{\beta^6}\right) \ . }
Note that the end result for the fundamental period of the subsystem does only depend on the invariant combination, $\beta^6$, with respect to the induced Greene-Plesser group.

We should remark that the structure of subsystems arising from the two divisors, $Q_1$ and $Q_2$, are rather different. As shown the periods of the subsystems are in both examples identified with the periods of K3 surfaces. This seems to a general feature for divisors, which just depend on a single algebraic modulus. However, for the former divisor, $Q_1$, the associated K3~surface is a function of two algebraic moduli, $\alpha(\psi, \xi)$ and $\phi$, whereas for the latter divisor, $Q_2$, the K3~surface varies only with respect to a single algebraic modulus, $\beta(\psi,\phi,\zeta)$, which is a combination of the two bulk and boundary algebraic moduli. A thorough treatment of the relevance of the K3~geometry will be given in ref.~\WProg.

%%%%%%%%%%%%%%%%%%%%%%
\subsec{Algebraic variables in the large complex structure regime}
%%%%%%%%%%%%%%%%%%%%%%
Before we discuss the remaining relative periods we first determine good algebraic variables in the large complex structure regime of the B-model, which allow to make the connection with the flat coordinates in the large radius regime of the A-model. As explained in detail in ref.~\AlimRF\ the good algebraic variables can be determined in the vicinity of the large complex regime by methods of toric geometry. Guided by their techniques we determine here these moduli space coordinates by imposing that all the periods should at most have logarithmic singularities at the large complex structure point and that the algebraic coordinates should be single valued with respect to the $\IZ_{12}$-group action discussed in section~\secBbraneGeom.

Then we obtain for the divisor, $Q_1$, the open-string algebraic variables
\eqn\AlgVarOpen{
  \hat z_1\,=\,\frac{\xi}{12\psi} \ , \quad
  \hat z_2\,=\,\frac{2\phi}{(12\psi)^3\xi^3} \ , \quad
  \hat z_3\,=\,\frac{1}{(2\phi)^2} \ . }
Note that these coordinates are invariant with respect to the $\IZ_{12}$~group action. Furthermore, as we will see, all relative periods expressed in terms of these coordinates have only logarithmic singularities. We also observe here that with respect to the $\IZ_2$~symmetry~\SymAlgQOne\ the new algebraic open-string coordinates transform as
\eqn\AlgSymOpen{\IZ_2\,:\ (\hat z_1,\hat z_2,\hat z_3)
  \,\mapsto\,\left(1-\hat z_1,\frac{\hat z_1^3\hat z_2}{(1-\hat z_1)^3},\hat z_3\right) \ . }
The deduced open-string coordinates are related to the closed string coordinates in ref.~\CandelasDM\ by
\eqn\AlgVarClosed{
  z_1\,=\,\hat z_1^3 \hat z_2\,=\,\frac{2\phi}{(12\psi)^6} \ ,\quad z_2\,=\,\hat z_3\,=\,\frac{1}{(2\phi)^3} \ . }
The subsytem fundamental period~\SubfundPerOne\ becomes in terms of these open-string coordinates
\eqn\SubFundPerLV{
   \chi_0(\hat z_1,\hat z_2,\hat z_3)\,=\,\frac{1}{12\,\psi\,\xi}\sum_{n=0}^{+\infty}
   \sum_{r=0}^{\left[\frac{n}{2}\right]}\frac{(3n)!}{(n!)^2(r!)^2(m-2r)!}
   \frac{\hat z_2^n\hat z_3^r}{(\hat z_1-1)^{3n+1}}  \ . }

In a similar way we determine the open-string coordinates corresponding to the large complex structure regime for the second divisor, $Q_2$, which are given by
\eqn\AlgSymOpenTwo{
   \hat w_1\,=\,\frac{2\phi}{\zeta} \ , \quad \hat w_2\,=\,\frac{\zeta}{(12\psi)^6} \ , \quad \hat w_3\,=\,\frac{1}{(2\phi)^2}  \ . }
Here the relation to the closed-string coordinates, $z_1$ and $z_2$, reads
\eqn\AlgVarClosedTwo{
  z_1\,=\,\hat w_1 \hat w_2 \ ,\quad z_2\,=\,\hat w_3 \ . }
%

%%%%%%%%%%%%%%%%%%%%%%
\subsec{Relative periods, open-string flat coordinates and the open-string mirror map}
\subseclab\secFlatCoord
%%%%%%%%%%%%%%%%%%%%%%
In order to obtain the remaining relative periods we examine the analytic properties of the fundamental periods. Here we review the performed analysis schematically. For more details we refer the interested reader to ref.~\CandelasDM.

In order to derive the mirror map we need to determine three independent logarithmic solutions of the open-/closed Picard-Fuchs system. By the general structure of the solutions to the open-/closed Picard-Fuchs operators \JockersPE, we know that the bulk periods of the closed-string sector appear also as relative periods in the open-/closed deformation space. These periods can be obtained by studying the analytic behavior of the fundamental period in terms of a Barnes integral and one arrives for the logarithmic solutions at \CandelasDM
\eqn\LogPerLR{\eqalign{
  \Pi^{\log z_1}&(z_1,z_2)\,=\,
      \Pi_0(z_1,z_2)\left(\log z_1-\sum_{n=1}^{+\infty}\frac{(2n-1)!}{(n!)^2}z_2^n\right)
      +\sum_{n=1}^{+\infty}\frac{(6n)!(-1)^n}{(n!)^3(3n)!}z_1^n \cr
      &\times\left(\sum_{r=0}^{\left[\frac{n}{2}\right]}\frac{n!\,z_2^r}{r!(n-2r)!}
      \left(6 \Psi(6n+1)-3\Psi(3n+1)-2\Psi(n+1)-\Psi(k+1)\right)-\frac{\hat f_n}{2} \right)  \ , \cr
  \Pi^{\log z_2}&(z_1,z_2)\,=\,\Pi_0(z_1,z_2)\left(\log z_2-\sum_{n=1}^{+\infty}\frac{(2n-1)!}{(n!)^2}z_2^n\right)
      -\sum_{n=1}^{+\infty}\frac{(6n)!(-1)^n}{(n!)^3(3n)!}z_1^n\hat f_n \ , }}
where the functions, $\hat f_n$, are defined recursively by \CandelasDM
\eqn\fhat{\hat f_n(\phi)\,=\,-\frac{\sqrt{\phi^2-1}}{(2\phi)^n} f_n(\phi) \ . }
with
\eqn\frec{f_n(\phi)\,=\,\frac{2(2n-1)\phi}{n} f_{n-1}(\phi)-\frac{4(n-1)(\phi^2-1)}{n}f_{n-2}(\phi) \ . }

However, in order to state the open-/closed-string mirror map we need in addition to the above logarithmic periods another linearly independent logarithmic solution to the Picard-Fuchs system. The remaining logarithmic period is obtained by integrating the fundamental period of the subsystem. For instance for the divisor, $Q_1$, we obtain from the fundamental period, $\chi_0$, through the relation, $\partial_\xi\rel\Pi=\chi_0$, another linearly independent logarithmic solution, which is given by
\eqn\FirstLogPerLR{\eqalign{
   \rel\Pi_1(\hat z_1,\hat z_2,\hat z_3)&\,=\,\rel\Pi_0(\hat z_1,\hat z_2,\hat z_3)\big(\log\hat z_1-\log(1-\hat z_1)\big)\cr
      +&2\sum_{m=1}^{+\infty}\sum_{n=0}^{3m-1}\sum_{r=0}^{\left[\frac{m}{2}\right]}
        \frac{(3m-n-1)!\left(3m-n+\frac{1}{2}\right)_n}{(m!)^2 (r!)^2 (m-2r)!}
        \frac{(4\hat z_1)^n(-\hat z_2)^m \hat z_3^r}{1-\hat z_1}\left(\hat z_1-\frac{1}{2}\right) \ . }}
Note that in the integration process there arises in general an integration constant depending only on the closed-string parameters, $z_1$ and $z_2$. This integration constant can be fixed either by using the Picard-Fuchs equations or by  studying the analytic behavior of the resulting relative period~\FirstLogPerLR.

Here, however, we can also use the $\IZ_2$~symmetry~\AlgSymOpen\ to determine the correct integration constant. By construction the Picard-Fuchs equations, which are obtained from the relative residue integral~\ResTwoFormCY, obey the $\IZ_2$~symmetry. As a consequence the relative periods of the open-/closed-deformation space can be separated into either even or odd relative periods. Since the subsystem periods are even with respect to this $\IZ_2$~symmetry and since the closed string algebraic coordinates, $z_1$ and $z_2$, are invariant, the integrated relative period~\FirstLogPerLR\ can be chosen to be odd, which fixes the integration constant unambiguously. The obtained integrated period~\FirstLogPerLR\ is manifestly odd under the $\IZ_2$~symmetry, and hence it is a solution to the Picard-Fuchs differential operators of the whole system.

Together with eqs.~\LogPerLR\ we now obtain the logarithmic relative periods for the open-string variables, $\hat z_2$ and $\hat z_3$,
\eqn\OtherLogPerLR{\eqalign{
   \rel\Pi_2(\hat z_1,\hat z_2,\hat z_3)&\,=\, \Pi^{\log z_1}(\hat z_1^3\hat z_2,\hat z_3)-3\,\rel\Pi_1(\hat z_1,\hat z_2,\hat z_3)  \ , \cr
   \rel\Pi_3(\hat z_1,\hat z_2,\hat z_3)&\,=\,  \Pi^{\log z_2}(\hat z_1^3\hat z_2,\hat z_3) \ , }}
These periods allow us to define in the usual way the open-string mirror coordinates, $\hat t_\ell$, via
\eqn\AModelCoord{2\pi i\,\hat t_\ell \,=\, \frac{\rel\Pi_\ell(\hat z_1,\hat z_2, \hat z_3)}{\rel\Pi_0(\hat z_1,\hat z_2, \hat z_3)}
   \,=\,\log \hat z_\ell + O(\hat z_\ell) \ , \quad \ell\,=\,1,2,3 \ . }
The K\"ahler coordinate, $\hat t_3$, is the coordinate associated to the cohomology class, $\omega_2$, whereas the coordinates, $\hat t_1$ and $\hat t_2$, are the volumes of two homologically distinct disks ending on the described family of A-branes. Note that the volume associated to the spheres of the cohomology class, $\omega_1$, are measured by $3\,\hat t_1+\hat t_2$.

From the definition of the K\"ahler coordinates we obtain recursively to leading order the open-string mirror map
\eqn\OpenFlatCoord{\eqalign{
    \hat z_1\,&=\,\hat q_1-\hat q_1^2+\hat q_1^3-2\hat q_1\hat q_2-14\hat q_1^2\hat q_2
          +5\hat q_1\hat q_2^2-2\hat q_1\hat q_2\hat q_3+\ldots \ , \cr
    \hat z_2\,&=\,\hat q_2+3\hat q_1\hat q_2+3\hat q_1^2\hat q_2+6\hat q_2^2+66\hat q_1\hat q_2^2
          +9\hat q_2^3+\hat q_2 \hat q_3+3 \hat q_1 \hat q_2 \hat q_3+12 \hat q_2^2 \hat q_3+\ldots \ , \cr
    \hat z_3\,&=\,\hat q_3-2\hat q_3^2+3\hat q_3^3+\ldots \ , }}
with $\hat q_\ell=e^{2\pi i \hat t_\ell},\ \ell=1,2,3$.

For the second divisor, $Q_2$, the remaining logarithmic solutions associated to the integrated subsystem reads
\eqn\FirstLogLRTwo{\eqalign{\rel\Pi_1(\hat w_1, \hat w_2, \hat w_3)\,&=\,-\rel\Pi_0(\hat w_1, \hat w_2, \hat w_3)\log(\hat{w}_{1})\cr
& +\sum_{{k=0\atop {k\neq m+n}}}^{\infty}\sum_{n=0}^{k}\sum_{m=0}^{n}\frac{(-1)^{n-m}(6k)!}{(k!)^{2}(3k)!(k-n)!(n-m)!m!}\,
\frac{1}{k-n-m}\,\hat{w}_{1}^{m+n}\hat{w}_{2}^{k}\hat{w}_{3}^{m}\cr
&+C(\hat{w}_{1}\hat{w}_{2},\hat{w}_{3}) \ . }}
Here the function, $C$, is the constant of integration, which by construction can only depend on the bulk large volume algebraic coordinates, $z_1=\hat w_1\hat w_2$ and $z_2=\hat w_3$. The leading order terms of the integration constant turn out to be
\eqn\consinteg{
 C(z_{1},z_{2})=\,z_{2}+\frac{3}{2}\,z_{2}^{2}-120\,z_{1}+\frac{10}{3}\,z_{2}^{3}+120\,z_{1}z_{2}
+\frac{35}{4}\,z_{2}^{4}+60\,z_{1}z_{2}^{2}+124\,740\,z_1^{2}+\cdots\ .}
whereas the remaining logarithmic solutions are given by
\eqn\OtherLogPerLRTwo{\eqalign{
   \rel\Pi_2(\hat w_1, \hat w_2, \hat w_3)\,&=\, \Pi^{\log z_1}(\hat w_1\hat w_2,\hat w_3)-\rel\Pi_1(\hat w_1,\hat w_2,\hat w_3) \ , \cr
   \rel\Pi_3(\hat w_1, \hat w_2, \hat w_3)\,&=\, \Pi^{\log z_2}(\hat w_2\hat w_2,\hat w_3) \ . }}
From these logarithmic periods we obtain the large radius leading order behavior
\eqn\OpenFlatCoordTwo{\eqalign{
 \hat w_1\,&=\,\hat{q}_{1}+120\hat{q}_{1}\hat{q}_{2}+\hat{q}_{1}\hat{q}_{3}+120\hat{q}_{1}\hat{q}_{2}\hat{q}_{3}-
 120\hat{q}_{1}^{2}\hat{q}_{2}+34380\hat{q}_{1}\hat{q}_{2}^{2}+\cdots \ , \cr
 \hat w_2\,&=\, \hat q_2 -120\hat{q}_{2}^{2}+864\hat{q}_{1}\hat{q}_{2}^{2}-19980\hat{q}_{2}^{3}
 +\cdots\ , \cr
 \hat w_3\,&=\, \hat q_3 -2\hat{q}_{3}^{2}+3\hat{q}_{3}^{3}+240\hat{q}_{1}\hat{q}_{2}\hat{q}_{3}
 +\cdots \ . }}
%

%%%%%%%%%%%%%%%%%%%%%%
\newsec{$N=1$ superpotentials and disk instantons}
%%%%%%%%%%%%%%%%%%%%%%
In this section we extract from the relative periods candidates for instanton generated A-brane $N=1$ superpotentials. However, in order to really interpret the stated numbers as disk Gromov-Witten invariants, a more detailed analysis of the classical terms together with a precise definition of the A-model geometry is necessary \WProg. Independent of the physical and geometric interpretation of the discussed integer invariants encoded in the computed $N=1$ superpotentials, it should be possible to proof their integrality by similar arguments as presented in ref.~\SchwarzTI.

%%%%%%%%%%%%%%%%%%%%%%
\subsec{Double-logarithmic relative periods}
%%%%%%%%%%%%%%%%%%%%%%
In order to construct the candidate integral linear combinations for the $N=1$ superpotential periods, we construct first all double-logarithmic solutions of the Picard-Fuchs differential equations. By employing the analytic technique or by solving the bulk Picard Fuchs equations of ref.~\CandelasDM\ one obtains for the double-logarithmic bulk periods the expressions
\eqn\BulkDoubleLog{\eqalign{
  \Pi^{\log^2z_1}(z_1,z_2)\,&=\,\Pi_0(z_1,z_2)(\log z_1)^2+2 A^{\log z_1}(z_1,z_2)\log z_1
  +C^{\log^2z_1}(z_1,z_2) \ ,  \cr
  \Pi^{\log z_1\log z_2}(z_1,z_2)\,&=\,\Pi_0(z_1,z_2)\log z_1\log z_2+A^{\log z_2}(z_1,z_2)\log z_1\cr
  &\quad\quad+A^{\log z_1}(z_1,z_2)\log z_2+C^{\log z_1\log z_2}(z_1,z_2) \ , }}
in terms of the constants
\eqn\BulkDoubleConst{\eqalign{
   C^{\log^2z_1}(z_1,z_2)\,&=\,2\,z_2+553\,536\, z_1^2-1\,728\,z_1z_2+{11\over 2}z_2^2+\ldots \ , \cr
   C^{\log z_1\log z_2}(z_1,z_2)\,&=\,-1248\,z_1-2\,z_2+1\,480\,896\,z_1^2+1\,978\,z_1z_2-{13\over 2}z_2^2+\ldots  \ , }}
where the functions, $A^{\log z_1}(z_1, z_2)$ and $A^{\log z_2}(z_1,z_2)$, are the non-singular parts in the vicinity, $z_1=z_2=0$, of the respective logarithmic bulk periods~\LogPerLR\ at the large radius regime.

Next we focus on the brane geometry associated to the divisor, $Q_1$, and construct the double-logarithmic relative periods, which are intrinsic to the brane geometry. Analogously as sketched in section~\secFlatCoord\ for the bulk periods, by examining the analytic properties of the fundamental period of the K3~geometry~\HyperPolSub\ of the subsystem, we first generate the two logarithmic periods, $\chi_1$ and $\chi_2$, of the subsystem.
\eqn\SubLogPerOne{\eqalign{
  \chi_1(\alpha,\phi)\,&=\,\chi_0(\alpha,\phi)\log\alpha^{-3} \cr &\quad +
     \sum_{n=1}^{+\infty}\sum_{r=0}^{\left[\frac{n}{2}\right]} \frac{(3n)!(-1)^n(2\phi)^{n-2r}}{(n!)^2(r!)^2(n-2r)!\alpha^{3n}}
       \left(3\Psi(3n+1)-2\Psi(n+1)-\Psi(k+1)\right) \ , \cr
  \chi_2(\alpha,\phi)\,&=\,\chi_0(\alpha,\phi)\log\left(\frac{\phi^{-2}}{\left(1+\sqrt{1-\phi^{-2}}\right)^2}\right)
      -\sqrt{\phi^2-1} \sum_{n=1}^{+\infty}\frac{(3n)!(-1)^n}{(n!)^3(\alpha)^{3n}}f_n(\phi) \ . }}
Here the functions, $f$, is recursively defined in eq.~\frec. By integrating these two logarithmic solutions of the subsystem, we finally arrive at the double-logarithmic relative periods associated to the divisor, $Q_1$. As explained in Section~\secFlatCoord, we determine again the integration constant by requiring odd parity with respect to the $\IZ_2$~symmetry. The result of this analysis yields
\eqn\SuperPer{\eqalign{
  \rel\Pi_4(\hat z_1,\hat z_2,\hat z_3)\,&=\,
      \frac{1}{2}\rel\Pi_0(\hat z_1,\hat z_2,\hat z_3)\log\hat z_1
      \left(3\,\log\hat z_1+2\,\log\hat z_2\right)\cr
      &\quad +\frac{A_3(\hat z_1,\hat z_2,\hat z_3)}{2}\log\hat z_1
      +\frac{A_1(\hat z_1,\hat z_2,\hat z_3)}{2}\log\hat z_3+C_4(\hat z_1,\hat z_2,\hat z_3) \ , \cr
  \rel\Pi_5(\hat z_2,\hat z_2,\hat z_3)\,&=\,
       \frac{1}{2}\rel\Pi_0(\hat z_1,\hat z_2,\hat z_3)\log\hat z_1\log\hat z_3 \cr
       &\quad +\frac{A_3(\hat z_1,\hat z_2,\hat z_3)}{2}\log\hat z_1
       +\frac{A_1(\hat z_1,\hat z_2,\hat z_3)}{2}\log\hat z_3+C_5(\hat z_1,\hat z_2,\hat z_3)  \ .}}
where the functions, $A_1(\hat z_1,\hat z_2,\hat z_3)$, $A_2(\hat z_1,\hat z_2,\hat z_3)$ and $A_3(\hat z_1,\hat z_2,\hat z_3)$, are the non-singular pieces of the logarithmic relative periods~\FirstLogPerLR\ and \OtherLogPerLR\ in the vicinity, $\hat z_1=\hat z_2=\hat z_3=0$. Furthermore, the leading terms of the non-singular series, $C_4(\hat z_1,\hat z_2,\hat z_3)$ and $C_5(\hat z_1,\hat z_2,\hat z_3)$, are given by
\eqn\CPolys{\eqalign{
  C_4(\hat z_1,\hat z_2,\hat z_3)\,&=\,
   6 \hat{z}_1+3 \hat z_2+3 \hat z_1^2+21 \hat z_1 \hat z_2-\frac{141}{4} \hat z_2^2-\hat z_1 \hat z_3+2 \hat z_2 \hat z_3+\ldots \ ,\cr
  C_5(\hat z_1,\hat z_2,\hat z_3)\,&=\,4 \hat z_2+24 \hat z_1 \hat z_2-45 \hat z_2^2+2 \hat z_1 \hat z_3-4 \hat z_2 \hat z_3+\ldots  \ .}}

In a similar way we generate the double logarithmic relative period for the family of branes associated to the second divisor, $Q_2$. Since the associated K3~geometry of the subsystem is parametrized by a single modulus the analysis is even simpler as there is only a single logarithmic subsystem period, $\vartheta_1$,
\eqn\SubLogPerTwo{\eqalign{ \vartheta_1(\beta)\,=&\,\vartheta_0(\beta)\log\left(\frac{1}{2^{6}\cdot 3^{3}\beta^{6}}\right)\cr
&+3\sum_{k=0}^{\infty}
\frac{(6k)!}{(k!)^{3}(3k)!}\Big(2\Psi(6k+1)-\Psi(3k+1)-\Psi(k+1)\Big)\frac{1}{(2^{6}\cdot 3^{3}\beta^{6})^{k}} \ , }}
which yields after integration the double-logarithmic relative period
\eqn\SuperPer{\eqalign{ \rel\Pi_4(\hat w_1,\hat w_2,\hat w_3)\,=&\,\,\rel\Pi_{0}(\hat{w}_{1},\hat{w}_{2},\hat{w}_{3})\log\hat{w}_{1}\Big(\log\hat{w}_{2}+\frac{1}{2}
\log\hat{w}_{3}\Big)\cr
&+A_{1}(\hat{w}_{1},\hat{w}_{2},\hat{w}_{3})\log(\hat{w}_{1}\hat{w}_{2})+A_{2}(\hat{w}_{1},\hat{w}_{2},\hat{w}_{3})
\log\hat{w}_{1}+C_{4}(\hat{w}_{1},\hat{w}_{2},\hat{w}_{3})  \ , }}
where the functions, $A_1( \hat{w}_{1},\hat{w}_{2},\hat{w}_{3} )$ and $A_2(\hat{w}_{1},\hat{w}_{2},\hat{w}_{3} )$, are the non-singular pieces of the logarithmic relative periods~\FirstLogLRTwo\ and \OtherLogPerLRTwo\ in the vicinity of the large radius point, $\hat w_1=\hat w_2=\hat w_3=0$. Furthermore, the function, $C_{4}(\hat{w}_{1},\hat{w}_{2},\hat{w}_{3})$, enjoys the expansion
\eqn\constfourB{C_{4}(\hat{w}_{1},\hat{w}_{2},\hat{w}_{3})=12\,\hat{w}_{3}-7488\,\hat{w}_{2}-12\,\hat{w}_{1}
+33\,\hat{w}_{3}^{2}-3128112\,\hat{w}_{2}^{2}+8928\,\hat{w}_{1}\hat{w}_{2}-3\,\hat{w}_{1}^{2}+\cdots\ .}
%

%%%%%%%%%%%%%%%%%%%%%%
\subsec{$N=1$ superpotentials and disk instantons}
%%%%%%%%%%%%%%%%%%%%%%
In the context of Calabi-Yau fourfolds the $N=1$ flux-induced large volume superpotentials are encoded in the double-logarithmic periods \MayrSH. There the correct linear combination of double-logarithmic periods is determined by matching the leading classical terms with the topological intersection numbers of the A-model geometry.

For brane geometries in Calabi-Yau threefolds the $N=1$ D-brane superpotentials of the A-model are expected to be purely disk instanton generated \refs{\OoguriCK,\KachruIH}, and hence we should not expect any classical terms at all. By construction, however, the (double) logarithmic periods give always rise to apparent classical terms. The role of these leading order apparent classical terms is not clear to us at the moment. Therefore, our strategy is to construct brane superpotential candidates from integral double-logarithmic relative periods and assume that there exists a suitable A-brane configuration on the mirror side for each of the stated superpotentials. We postpone the necessary general treatment to ref.~\WProg.

The quantum piece, $W_{\rm Disk}$, in the constructed superpotentials encode via the multi-covering formula \refs{\OoguriBV,\LabastidaZP}
\eqn\DiskMC{
  W_{\rm Disk}({\vec t})\,=\,\sum_{\vec n}\sum_{k=1}^{+\infty} N_{\vec n}
  \frac{e^{2\pi i\,{\vec t}\cdot{\vec n}}}{k^2} \ , }
integral invariants, $N_{\vec n}$. Hence, we expect that, after constructing the associated Lagrangian submanifold in the A-model, these integral invariants can be given a geometric interpretation as disk Gromov-Witten invariants, $N_{\vec n}$. Then the vector, $\vec t$, encodes the open-string K\"ahler moduli, $\hat t_\ell$, measuring the volumes of homologcially distinct disks, whereas the multi-index, $\vec n$, labels the different integral Gromov-Witten disk invariants, $N_{\vec n}$ \AlimRF.

Presumably, it is possible to show in full generality the integrality property of the integral invariants independent from an interpration as disk worldsheet instantons \SchwarzTI. However, the integrality property of the invariants does not automatically guaranty the existence of a suitable A-model brane geometry. At the moment we are not aware of a reliable criterion to ensure the anticipated physical interpretation of all these integral numbers.

Let us now come back to our explicit example. First we analyze the family of branes captured by the divisor, $Q_1$. By requiring integrality we can extract two $N=1$ brane superpotential candidates. This is a consequence of the fact that the subsystem K3~geometry depends on two algebraic moduli. After inserting the open-string mirror map~\OpenFlatCoord\ the associated superpotentials are explicitly given by
\eqn\SuperPotOne{\eqalign{
  W^{(A)}(\hat t_1,\hat t_2,\hat t_3)\,&=\,\frac{1}{(2\pi i)^2}
    \frac{\rel\Pi_4(\hat z_1(\hat t),\hat z_2(\hat t),\hat z_3(t))}{\rel\Pi_0(\hat z_1(\hat t),\hat z_2(\hat t),\hat z_3(t))}
    \,=\, \hat t_1\hat t_2+\frac{3}{2}\hat t_1^2+\frac{1}{(2\pi i)^2}W^{(A)}_{\rm Disk}(\hat q_1,\hat q_2,\hat q_3) \ , \cr
  W^{(B)}(\hat t_1,\hat t_2,\hat t_3)\,&=\,\frac{1}{(2\pi i)^2}
    \frac{\rel\Pi_5(\hat z_1(\hat t),\hat z_2(\hat t),\hat z_3(t))}{\rel\Pi_0(\hat z_1(\hat t),\hat z_2(\hat t),\hat z_3(t))} 
    \,=\, \frac{1}{2}\hat t_1\hat t_3+\frac{1}{(2\pi i)^2}W^{(B)}_{\rm Disk}(\hat q_1,\hat q_2,\hat q_3) \ . }}
The quantum pieces, $W_{\rm Disk}^{(A)}$ and $W_{\rm Disk}^{(B)}$, have the predicted integral structure encoded in the multi-covering formula~\DiskMC. The first few integral invariants are recorded in \ltab\NWA\ and \ltab\NWB, respectively.

Note that the discrete $\IZ_2$~symmetry~\AlgSymOpen\ gives rise to the $\IZ_2$~quantum symmetry on the mirror A-model superpotentials, namely the instanton numbers obey
\eqn\QSym{\IZ_2:\ N_{n,k,l} \mapsto - N_{6k-n,k,l} \ , }
which corresponds on the level of the open-string moduli fields, $\hat t$, to the symmetry operation
\eqn\QSymt{\IZ_2:\ (\hat t_1, \hat t_2, \hat t_3) \mapsto (-\hat t_1, \hat t_2+6\,\hat t_1, \hat t_3) \ . }
Note that for the superpotential, $W^{(A)}$, the classical terms are not invariant with respect to the quantum symmetry by themselves. Instead the instanton numbers, $N^{(A)}_{1,0,0}=6$ and $N^{(A)}_{2,0,0}=-3$, which naively are mapped to $N^{(A)}_{-1,0,0}$ and $N^{(A)}_{-2,0,0}$, give rise to a resummation of the instanton series, which restores the invariance of the whole superpotential, $W^{(A)}$.

In addition we observe that the instanton numbers of the superpotential, $W^{(B)}$, in \ltab\NWB\ exhibit another quantum symmetry
\eqn\QSum{\IZ_2:\ N_{n,k,l} \mapsto - N_{n,k,k-l} \ , }
which is the open-string analog of the closed-string instanton quantum symmetry, $n_{s,t}=n_{s,s-t}$, discussed in ref.~\CandelasDM. 

%%%%%%%%%%%%%%%%%%%%%%%%%%%%%%%%%%
\tabinsert\NWA{The table lists some of the integral invariants encoded in the superpotential, $W^{(A)}$, arising from the divisor, $Q_1$.}
{\centerline{\vbox{
\offinterlineskip
\tabskip=0pt\halign{
\vrule height10pt depth6pt#\tabskip=2.5pt plus 1fil\strut
   &\hfil#\hfil&\vrule#&\hfil#\hfil&\hfil#\hfil&\hfil#\hfil&\hfil#\hfil&\vrule width1.5pt#&\hfil#\hfil&\vrule#&
   \hfil#\hfil&\hfil#\hfil&\hfil#\hfil&\hfil#\hfil&\tabskip=0pt\vrule#\cr
\noalign{\hrule}
&$\scriptstyle N^{(A)}_{n,k,0}$ &&
$\ \ k=0\ \ $ & $\ \quad1\quad\ $ & $\ \quad2\quad\ $ & $\ \quad3\quad\ $ && 
$\scriptstyle N^{(A)}_{n,k,1}$ &&
$\ \ k=0\ \ $ & $\ \quad1\quad\ $ & $\ \quad2\quad\ $ & $\ \quad3\quad\ $ &\cr 
\noalign{\hrule height1pt}
&$\,n=0$ && $--$ & $3$ & $-12$ & $81$ && $\,n=0$ && $0$ & $7$ & $-129$ & $2356$&\cr
&$\phantom{\,n=\,}1$ && $6$ & $24$ & $-150$ & $1536$ && $\phantom{\,n=\,}1$ && $0$ & $60$ & $-1812$ & $46308$&\cr
&$\phantom{\,n=\,}2$ && $-3$ & $51$ & $-828$ & $12789$ && $\phantom{\,n=\,}2$ && $0$ & $231$ & $-11688$ & $422769$&\cr
&$\phantom{\,n=\,}3$ && $0$ & $0$ & $-2448$ & $62424$ && $\phantom{\,n=\,}3$ && $0$ & $0$ & $-44796$ & $2369184$&\cr
&$\phantom{\,n=\,}4$ && $0$ & $-51$ & $-4284$ & $200175$ && $\phantom{\,n=\,}4$ && $0$ & $-231$ & $-118617$ & $9172275$&\cr
\noalign{\hrule height1.5pt}
&$\scriptstyle N^{(A)}_{n,k,2}$ &&
$\ \ k=0\ \ $ & $\ \quad1\quad\ $ & $\ \quad2\quad\ $ & $\ \quad3\quad\ $ && 
$\scriptstyle N^{(A)}_{n,k,3}$ &&
$\ \ k=0\ \ $ & $\ \quad1\quad\ $ & $\ \quad2\quad\ $ & $\ \quad3\quad\ $ &\cr 
\noalign{\hrule height1pt}
&$\,n=0$ && $0$ & $0$ & $-34$ & $3172$ && $\,n=0$ && $0$ & $0$ & $0$ & $261$&\cr
&$\phantom{\,n=\,}1$ && $0$ & $0$ & $-438$ & $62580$ && $\phantom{\,n=\,}1$ && $0$ & $0$ & $0$ & $4992$&\cr
&$\phantom{\,n=\,}2$ && $0$ & $0$ & $-2556$ & $576165$ && $\phantom{\,n=\,}2$ && $0$ & $0$ & $0$ & $42489$&\cr
&$\phantom{\,n=\,}3$ && $0$ & $0$ & $-8592$ & $3273720$ && $\phantom{\,n=\,}3$ && $0$ & $0$ & $0$ & $215676$&\cr
&$\phantom{\,n=\,}4$ && $0$ & $0$ & $-20250$ & $12926751$ && $\phantom{\,n=\,}4$ && $0$ & $0$ & $0$ & $738051$&\cr
\noalign{\hrule height1pt}
}}}}
%%%%%%%%%%%%%%%%%%%%%%%%%%%%%%%%%%

%%%%%%%%%%%%%%%%%%%%%%%%%%%%%%%%%%
\tabinsert\NWB{The table lists some of the integral invariants encoded in the superpotential, $W^{(B)}$, arising from the divisor, $Q_1$.}
{\centerline{\vbox{
\offinterlineskip
\tabskip=0pt\halign{
\vrule height10pt depth6pt#\tabskip=2.5pt plus 1fil\strut
   &\hfil#\hfil&\vrule#&\hfil#\hfil&\hfil#\hfil&\hfil#\hfil&\hfil#\hfil&\vrule width1.5pt#&\hfil#\hfil&\vrule#&
   \hfil#\hfil&\hfil#\hfil&\hfil#\hfil&\hfil#\hfil&\tabskip=0pt\vrule#\cr
\noalign{\hrule}
&$\scriptstyle N^{(B)}_{n,k,0}$ &&
$\ \ k=0\ \ $ & $\ \quad1\quad\ $ & $\ \quad2\quad\ $ & $\ \quad3\quad\ $ && 
$\scriptstyle N^{(B)}_{n,k,1}$ &&
$\ \ k=0\ \ $ & $\ \quad1\quad\ $ & $\ \quad2\quad\ $ & $\ \quad3\quad\ $ &\cr 
\noalign{\hrule height1pt}
&$\,n=0$ && $--$ & $2$ & $-11$ & $90$ && $\,n=0$ && $0$ & $-2$ & $0$ & $408$&\cr
&$\phantom{\,n=\,}1$ && $0$ & $18$ & $-144$ & $1728$ && $\phantom{\,n=\,}1$ && $0$ & $-18$ & $0$ & $8136$&\cr
&$\phantom{\,n=\,}2$ && $0$ & $90$ & $-864$ & $14850$ && $\phantom{\,n=\,}2$ && $0$ & $-90$ & $0$ & $76698$&\cr
&$\phantom{\,n=\,}3$ && $0$ & $0$ & $-3072$ & $76626$ && $\phantom{\,n=\,}3$ && $0$ & $0$ & $0$ & $452268$&\cr
&$\phantom{\,n=\,}4$ && $0$ & $-90$ & $-7983$ & $268938$ && $\phantom{\,n=\,}4$ && $0$ & $90$ & $0$ & $1877238$&\cr
\noalign{\hrule height1.5pt}
&$\scriptstyle N^{(B)}_{n,k,2}$ &&
$\ \ k=0\ \ $ & $\ \quad1\quad\ $ & $\ \quad2\quad\ $ & $\ \quad3\quad\ $ && 
$\scriptstyle N^{(B)}_{n,k,3}$ &&
$\ \ k=0\ \ $ & $\ \quad1\quad\ $ & $\ \quad2\quad\ $ & $\ \quad3\quad\ $ &\cr 
\noalign{\hrule height1pt}
&$\,n=0$ && $0$ & $0$ & $11$ & $-408$ && $\,n=0$ && $0$ & $0$ & $0$ & $-90$&\cr
&$\phantom{\,n=\,}1$ && $0$ & $0$ & $144$ & $-8136$ && $\phantom{\,n=\,}1$ && $0$ & $0$ & $0$ & $-1728$&\cr
&$\phantom{\,n=\,}2$ && $0$ & $0$ & $864$ & $-76698$ && $\phantom{\,n=\,}2$ && $0$ & $0$ & $0$ & $-14850$&\cr
&$\phantom{\,n=\,}3$ && $0$ & $0$ & $3072$ & $-452268$ && $\phantom{\,n=\,}3$ && $0$ & $0$ & $0$ & $-76626$&\cr
&$\phantom{\,n=\,}4$ && $0$ & $0$ & $7983$ & $-1877238$ && $\phantom{\,n=\,}4$ && $0$ & $0$ & $0$ & $-268938$&\cr
\noalign{\hrule height1pt}
}}}}
%%%%%%%%%%%%%%%%%%%%%%%%%%%%%%%%%%
 
Analogously, we now analyze the double-logarithmic periods of the second divisor, $Q_2$. The subsystem K3~geometry depends only on one modulus. As a consequence there is only a single integral double-logarithmic relative period:
\eqn\SuperPotTwo{\eqalign{
  W(\hat t_1,\hat t_2,\hat t_3)\,&=\,\frac{1}{(2\pi i)^2}
    \frac{\rel\Pi_4(\hat w_1(\hat t),\hat w_2(\hat t),\hat w_3(t))}{\rel\Pi_0(\hat w_1(\hat t),\hat w_2(\hat t),\hat w_3(t))}\cr
    \,&=\, \hat t_1^2+2\hat t_1\hat t_2+\hat t_1\hat t_3+\hat t_2\hat t_3 +\frac{1}{(2\pi i)^2}W_{\rm Disk}(\hat q_1,\hat q_2,\hat q_3) \ . }}
The integral invariants encoded in the quantum piece, $W_{\rm Disk}$, are listed in \ltab\NWC.

%%%%%%%%%%%%%%%%%%%%%%%%%%%%%%%%%%
\tabinsert\NWC{The table lists some of the integral invariants encoded in the brane superpotential, $W$, arising from the divisor, $Q_2$.}
{\centerline{\vbox{
\offinterlineskip
\tabskip=0pt\halign{
\vrule height10pt depth6pt#\tabskip=2.5pt plus 1fil\strut
   &\hfil#\hfil&\vrule#&\hfil#\hfil&\hfil#\hfil&\hfil#\hfil&\hfil#\hfil&\vrule width1.5pt#&\hfil#\hfil&\vrule#&
   \hfil#\hfil&\hfil#\hfil&\hfil#\hfil&\hfil#\hfil&\tabskip=0pt\vrule#\cr
\noalign{\hrule}
&$\scriptstyle N_{n,k,0}$ &&
$\ \ k=0\ \ $ & $\ \quad1\quad\ $ & $\ \quad2\quad\ $ & $\ \quad3\quad\ $ && 
$\scriptstyle N_{n,k,1}$ &&
$\ \ k=0\ \ $ & $\ \quad1\quad\ $ & $\ \quad2\quad\ $ & $\ \quad3\quad\ $ &\cr 
\noalign{\hrule height1pt}
&$\,n=0$ && $--$ & $-624$ & $-178440$ & $-102780624$ && $\,n=0$ && $1$ & $0$ & $0$ & $0$&\cr
&$\phantom{\,n=\,}1$ && $-1$ & $624$ & $290628$ & $224298752$ && $\phantom{\,n=\,}1$ && $-1$ & $-624$ & $290628$ & $224298752$&\cr
&$\phantom{\,n=\,}2$ && $0$ & $0$ & $-112188$ & $-150496256$ && $\phantom{\,n=\,}2$ && $0$ & $624$ & $0$ & $-706643136$&\cr
&$\phantom{\,n=\,}3$ && $0$ & $0$ & $0$ & $28978128$ && $\phantom{\,n=\,}3$ && $0$ & $0$ & $-290628$ & $331848128$&\cr
&$\phantom{\,n=\,}4$ && $0$ & $0$ & $0$ & $0$ && $\phantom{\,n=\,}4$ && $0$ & $0$ & $0$ & $150496256$&\cr
\noalign{\hrule height1.5pt}
&$\scriptstyle N_{n,k,2}$ &&
$\ \ k=0\ \ $ & $\ \quad1\quad\ $ & $\ \quad2\quad\ $ & $\ \quad3\quad\ $ && 
$\scriptstyle N_{n,k,3}$ &&
$\ \ k=0\ \ $ & $\ \quad1\quad\ $ & $\ \quad2\quad\ $ & $\ \quad3\quad\ $ &\cr 
\noalign{\hrule height1pt}
&$\,n=0$ && $0$ & $0$ & $0$ & $0$ && $\,n=0$ && $0$ & $0$ & $0$ & $0$&\cr
&$\phantom{\,n=\,}1$ && $0$ & $0$ & $0$ & $0$ && $\phantom{\,n=\,}1$ && $0$ & $0$ & $0$ & $0$&\cr
&$\phantom{\,n=\,}2$ && $0$ & $0$ & $112188$ & $-150496256$ && $\phantom{\,n=\,}2$ && $0$ & $0$ & $0$ & $0$&\cr
&$\phantom{\,n=\,}3$ && $0$ & $0$ & $-290628$ & $-331848128$ && $\phantom{\,n=\,}3$ && $0$ & $0$ & $0$ & $-28978128$&\cr
&$\phantom{\,n=\,}4$ && $0$ & $0$ & $128440$ & $706643136$ && $\phantom{\,n=\,}4$ && $0$ & $0$ & $0$ & $150496256$&\cr
\noalign{\hrule height1pt}
}}}}
%%%%%%%%%%%%%%%%%%%%%%%%%%%%%%%%%%

We should remark that if the listed integral invariants in \ltab\NWA, \ltab\NWB\ and \ltab\NWC\ do indeed enjoy an interpretation as open-string Gromov-Witten invariants, there could be a mismatch by an overall integral multiple as we determine the overall normalizations of the corresponding superpotentials merely by requiring integrality. 

%%%%%%%%%%%%%%%%%%%%%%
\newsec{Conclusions}
%%%%%%%%%%%%%%%%%%%%%%
In this work we have demonstrated the method of direct integration to economically derive the relative periods for divisors in Calabi-Yau hypersurfaces. Although we have worked with explicit divisors in a particular Calabi-Yau threefold, the presented procedure is generic and can be applied for other Calabi-Yau hypersurfaces and complete intersections as well. So far the divisor, however, is restricted to depend on a single open-string algebraic modulus. Then the solutions to the open-string subsystem Picard-Fuchs differential equations contain a distinguished fundamental period as the open-string subsystem geometry is associated to a K3~surface. For more moduli, however, the subsystem geometry becomes more complicated \WProg, and it would be interesting to generalize the direct integration approach also to multi-open-string moduli examples.

For the analyzed geometry we have also examined the double-logarithmic periods in order to extract $N=1$ brane superpotentials, which encode open-string disk instantons in the semi-classical regime of the corresponding A-brane configuration in the Calabi-Yau manifold. We assume that the extracted double-logarithmic relative periods qualify for valid $N=1$ brane superpotentials in the large radius phase of a suitable brane configuration in the mirror Calabi-Yau manifold. We state the arising integral invariants for the associated $N=1$ brane superpotential candidate. 

At the moment we have a very crude way to determine superpotential periods. As pointed out in the main text the observation that double-logarithmic periods give rise to integral invariants seems to be a general feature of the large radius solutions to relative Picard Fuchs differential equations as proven for related geometric scenarios in ref.~\SchwarzTI. In order to really establish that the analyzed integer invariants actually allow for an interpretation as open Gromov-Witten invariants further checks are necessary. A thorough analysis of the structure of the relative periods is required in addition to a systematic construction of the A-model geometry \WProg. Ideally one would like to start the analysis with a given special Lagrangian submanifold with a classically unobstructed open-string modulus in a Calabi-Yau manifold, so as to calculate the quantum obstructed superpotential on the mirror B-model side, by applying, for instance, techniques presented here and in refs.~\refs{\JockersPE,\AlimRF}. An independent computation directly in the topological A-model would further support the described mirror symmetry recipe.

%%%%%%%%%%%%%%%%%%%%%%%
\goodbreak\bigskip\noindent
{\bf Acknowledgments}
\medskip\noindent
%%%%%%%%%%%%%%%%%%%%%%%
We are grateful to Peter Mayr for sharing his ideas and insights in discussions and many correspondences. We would like to thank Murad Alim, Micheal Hecht, Peter Mayr and Adrian Mertens for patiently explaining their work to us. We are also thankful to Mina Aganagic, Jun Li, Shamit Kachru and Albert Schwarz for comments and discussions. H.J. thanks the hospitality of the LMU Munich, where part of this work was completed. M.S. is supported by the Mellam Family Graduate Fellowship. This work is also supported by the Stanford Institute for Theoretical Physics and by the NSF~Grant~0244728.

%%%%%%%%%%%%%%%%%%%%%%%
%\goodbreak\bigskip
%%%%%%%%%%%%%%%%%%%%%%

%%%%%%%%%%%%%%%%%%%%%%
\listrefs
\end